\documentclass[aps,jcp,twocolumn,superscriptaddress,showpacs]{revtex4-1}
\usepackage{epsfig}
\usepackage{color}
\usepackage{graphics}
\usepackage{graphicx}
\usepackage{amsmath}
\usepackage[capitalize]{cleveref}
\usepackage[caption=false]{subfig}
\begin{document}

\title{Competitive Heterogeneous Nucleation onto a Microscopic Impurity in a Potts Model }

\author{Cletus C. Asuquo}
\affiliation{Department of Chemistry, University of Saskatchewan, SK, S7N 5C9, Canada}

\author{Danielle McArthur}
\affiliation{Department of Chemistry, University of Saskatchewan, SK, S7N 5C9, Canada}

\author{Richard K. Bowles}
\affiliation{Department of Chemistry, University of Saskatchewan, SK, S7N 5C9, Canada}
\email{richard.bowles@usask.ca}

\begin{abstract}
Many metastable systems can nucleate to multiple competing stable or intermediate metastable states. In this work, a Potts model, subject to external fields, is used to study the competitive nucleation of two phases attempting to grow on a microscopic impurity. Monte Carlo simulations are used to calculate the free energy surfaces for the system under different conditions, where the relative stability of the phases is adjusted by changing the interaction parameters, and the nucleation rates obtained using multicomponent transition state theory are compared with the rates measured using the survival probability method. We find that the two methods predict similar nucleation rates when the free energy barrier used in the transition state theory is defined as the work required to form a critical embryo from the metastable phase. An analysis of the free energy surfaces also reveals that the competition between the nucleating phases leads to an effective drying of the impurity which slows down the nucleation rate compared to the single phase case.
\end{abstract}


\maketitle

\section{Introduction}
Nucleation is the activated process that controls the kinetics of phase transitions, such as the condensation of a vapor or the freezing of a liquid to a crystal. Classical nucleation theory~\cite{Debenedetti:1996tf,Kelton:2010wj} (CNT), which describes nucleation in terms of the creation of small embryos of the new stable phase, focuses on the case where there is a single stable state so that there is only one possible transition. However, many phase transitions occur under conditions where the initial metastable phase can nucleate to more than one phase due to the presence of intermediate metastable phases. For example, molecular dynamic simulations~\cite{Chushak:2001p1372,Asuquo:2012p1373} and experiments~\cite{Koga:2004cs,Koga:2006p12920} have shown that liquid nanoclusters can freeze to a number of different structures including icosahedra and decahedra, even under conditions where the face centered cubic (FCC) base cluster is the most stable state. Bulk materials such as water and silica also exhibit polymorphism~\cite{Debenedetti:1996tf,Kelton:2010wj} and knowing how different crystal structures compete under a given set of conditions is essential for developing new materials~\cite{Eddaoudi:2002p1673} and pharmaceuticals~\cite{Morris:2001p1672}, and for understanding important atmospheric problems~\cite{Murray:2005fq,Murray:2006fn}.

Ostwald's step rule~\cite{Ostwald:1897wd} originally suggested that the metastable state will initially nucleate to the phase that is closest in terms of free energy, which will in turn nucleate to the next closest so that the system eventually ``steps" its way down to the most stable state, but it is now generally accepted that the lowest free energy barrier exiting the metastable region will determine which phase nucleates first~\cite{stranksi1933}. While a number of systems have been shown to follow the step rule~\cite{vanMeel:2008p8395,Chung:2009fe,Peng:2014is}, there is growing evidence that the presence of intermediate, or competing, metastable states can lead to a variety of nucleation pathways~\cite{Sear:2009it,Hedges:2011ci}. Ten Wolde and Frenkel~\cite{Wolde:1997ez} showed that the presence of a metastable fluid-fluid critical point could significantly alter the fluid-crystal nucleation mechanism and its nucleation rate even though the phase transition does not actually visit the free energy basin associated with the intermediate metastable state. Similar effects, involving the formation of precritical liquid clusters, have also been observed in a solid-solid phase transition of confined hard spheres~\cite{Qi:2015ie}. 


In this paper, we study the competitive nucleation of two phases attempting to grow on a single nanoscale impurity using a simple Potts model. The rate of nucleation to the different stable phases is calculated using two methods, the survival probability (SP)~\cite{Chushak:2001p1372}, and transition state theory (TST) in the form developed by Volmer and Weber~\cite{Volmer:1926p1353}, Becker and D{\"o}ring~\cite{Becker1935pg719}, Zeldovich~\cite{zeldovich1943}, and Frenkel~\cite{Frenkel:1959p1392}. The first method measures the rate of escape from the metastable phase by following an ensemble of molecular dynamics trajectories that end in nucleation. The rate at which each individual phase is formed is then obtained from the probabilities of observing the appearance of the phase as suggested by Sanders~\cite{Sanders:2007p994}. In contrast, the TST method focuses on the thermodynamic measurement of the free energy of forming a critical sized cluster for each phase. The methods are shown to predict nucleation rates within 50\% of each other over a series of different conditions and they both capture the same general features of competitive nucleation onto an impurity. In particular, we find that the overall rate of exiting the metastable state is slower in the case of competitive nucleation, compared to the non-competitive process, because the interactions between the competing phases reduce the wetting of the impurity. The remainder of the paper is organized as follows: Section II describes the model, Section III provides the details of the simulation methods used to calculate the nucleation rates, while our results and discussion are contained in Section IV. Section V contains our conclusions.

\section{Model}
\begin{figure}[t]
\begin{center}
\includegraphics[width=3.5in]{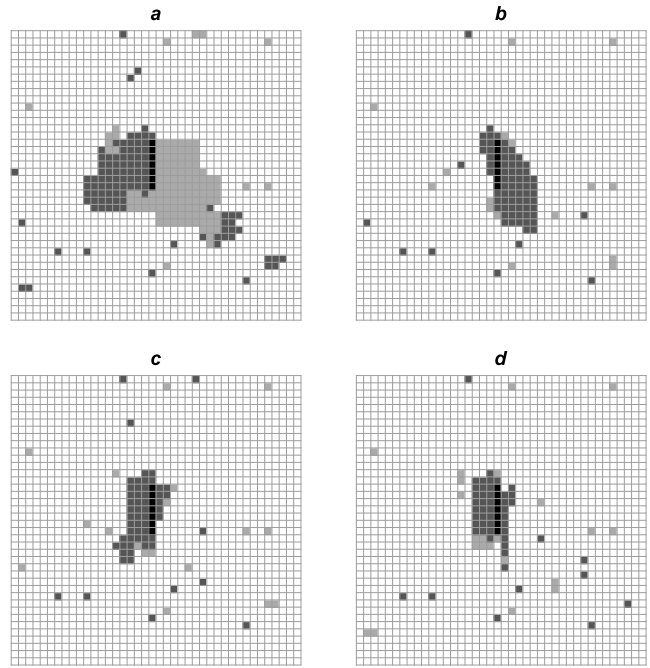}
\caption{(a) A  configuration of the Potts model for {\bf Case 1} containing a cluster growing on the impurity. The light grey and dark grey lattice sites represent the stable phases, $B$ and $C$, the black lattice sites represent the impurity and the white lattice sites represent the metastable phase, $A$. (b-d) Critical clusters for the $C$-transition in {\bf Cases 1-3}, respectively.}
\label{fig:pott}
\end{center}
\end{figure}

We study a four state Potts model~\cite{Wu:1982p1353,Arkin:2000p1125}, ($q=4$), where the first three spin states represent the metastable mother phase and the two competing more stable phases, labeled $A$, $B$ and $C$ respectively. The fourth state represents the heterogeneity, which consists of seven spins arranged in a line that are  located at the centre of the lattice and are unable to change state during the course of the simulation (See Fig.~\ref{fig:pott}). We have used a system of $N=L\times L=40\times40$ spins on a square lattice and have employed periodic boundaries. For a given configuration of the system, the energy is given by,
\begin{equation}
E(\sigma)=-\sum_{\left< i,j\right>}\textbf{J}_{\sigma_i,\sigma_j}-\sum_{\alpha=1}^qh_{\alpha}M_{\alpha}      \mbox{,}
\label{eq1}
\end{equation}
where $\textbf{J}_{\sigma_i,\sigma_j}$ is the interaction energy between nearest neighbour $\left<i,j\right>$ spins $\sigma_i$ and $\sigma_j$, $h_{\alpha}$ is the external field strength, which controls the relative stability of each of phase, and
\begin{equation}
M_{\alpha}=\sum_{i=1}^N\delta_{\sigma_i,\alpha}          \mbox{,}
\end{equation}
is the magnetization of spin type $\alpha$, where $\delta_{\sigma_i\alpha}=1$ if $\sigma_i=\alpha$ and 0 otherwise. The diagonal elements of  $\textbf{J}$ describe the interaction between spins of the same phase and setting $\textbf{J}_{\alpha,\alpha}=1.0$, for all $\alpha$, establishes the energy scale for the model. We also ensure all phases have the same favourable interaction with the heterogeneity, $\textbf{J}_{\alpha,4}=1.0$. With the temperature, $T$, fixed so that $k_BT=1.5$ and $h_A=-0.12$ for all simulations, we study three cases:
\begin{description}
\item[Case 1] $\textbf{J}_{B,C}=-1.0$, $\textbf{J}_{A,B}=\textbf{J}_{A,C}=0$ and $h_B= h_C=0.12$ so that phases $B$ and $C$ are equally stable relative to the mother phase, but have a strong dislike for each other.
\item[Case 2] $\textbf{J}_{B,C}=-1.0$, $\textbf{J}_{A,B}=\textbf{J}_{A,C}=0$, $h_B=0.12$ and $ h_C=0.17$ causes phase $C$ to become more stable.
\item[Case 3] $\textbf{J}_{B,C}=-0.8$, $\textbf{J}_{A,B}=\textbf{J}_{A,C}=0$, $ h_B=0.12$ and $ h_C=0.17$. By reducing the unfavourable interaction between the two stable phases, we allow a greater degree of mixing.
\end{description}
Finally, we also study aspects of heterogeneous nucleation in the Ising model~\cite{Scheifele:2013uo}, where $q=3$ so there are three phases, the metastable mother phase, a single stable phase and the impurity. In this case, the model parameters are selected as $\textbf{J}_{A,B}=0$ and $h_B=0.12$, to be consistent with {\bf Case 1}, which allows us to make a direct comparison between competitive and non-competitive nucleation processes.

\section{Methods}

\subsection{Cluster Criteria}
Both methods used here require a cluster definition to describe the growing nucleus. We follow Scheifele $et$ $al$~\cite{Scheifele:2013uo} and identify a cluster as the contiguous set of $B$ and $C$ spins that contain the impurity, so the cluster is characterized by the number of each type of spin, $(n_B,n_C)$ and the cluster size, $N_{c}=n_B+n_C$. Defined as such, a set of spins with small number of spins contacting one side of the impurity and a separate set of spins contacting the other side is considered to be a single cluster. This does not present any problems in counting the number of clusters as long as only one ``cluster" associated with the impurity goes over the nucleation barrier. The critical cluster sizes observed here are large in comparison to the impurity, and they completely wet the surface. Furthermore, the criteria identifies phase separated clusters, such as the one appearing in Fig.~\ref{fig:pott}, as a single cluster. 


\subsection{Survival Probability}
The survival probability has been used to study nucleation in a variety of systems, including the freezing of gold nanoparticles~\cite{Chushak:2001p1372,Asuquo:2012p1373}, crystallization in Lennard-Jones clusters~\cite{Lundrigan:2009p8124}, and crystallization in high pressure simulated silica~\cite{SaikaVoivod:2006p6}. Assuming first order kinetics, the nucleation rate, $J^{SP}$, for the system escaping the metastable state, is obtain from \cite{Chushak:2001p1372}
\begin{equation}
\text{ln}R(t)=-J^{SP}V(t-t_0)           \mbox{,}
\label{eq:esp}
\end{equation}
where $R(t)$ is the fraction of unnucleated systems at a given time, $t$, $V$ is the volume of the system and $t_0$ is the lag time.  In a bulk, uniform system, the volume term accounts for the translational degrees of freedom of the nucleating embryo because a nucleation event can occur anywhere with equal probability. In the presence of the microscopic impurity, only a single nucleation event can occur, located at the heterogeneity, so we set $V=1$. The rate is then reported in units of the number of clusters per unit time (mcs$^{-1}$) and the slope equals the overall rate at which the system leaves the metastable phase to form one of the stable phases.

To evaluate $R(t)$, we run 2000 independent trajectories using starting configurations with $n_B\sim0$, $n_C\sim0$. The trajectory is evolved using the standard metropolis Monte Carlo (MC) algorithm~\cite{Frenkel2002} where a randomly selected spin can be flipped to either of the other two phases with equal probability. The test MC move is then accepted or rejected according to the usual Boltzmann weighted probabilities for the change in energy. The unit of time is taken to be $N$ MC attempts so that on average each spin has the possibility of changing state in a single time step. The simulation is stopped when the largest cluster is greater than $N_{c}=60\%$ of the system. We determine that a system has nucleated to a given phase at a time when $n_i>150$, where $i=B$ or $C$, and $150$ should be much larger than the critical nucleus size (See Table~\ref{tab:summary} for actual critical sizes obtain from our free energy calculations).  This also allows us to determine the probability that a phase nucleates, $P_{\kappa}$, as the fraction of the total number of trajectories that end in a given phase so that the nucleation rate of a phase is given by~\cite{Sanders:2007p994},
\begin{equation}
J^{SP}_{\kappa}=J^{SP}P_{\kappa}\mbox{.}\\
\label{eq:irate}
\end{equation}

\subsection{Multicomponent Transition State Theory}
The two component nature of the clusters in our model means that the free energy surface describing the formation of a cluster is two dimensional and it will be necessary to describe nucleation in terms of the flux moving through a saddle point region. This problem has been studied in detail by Trinkaus~\cite{Trinkaus:1983eg} and Wilemski ~\cite{Wilemski:1999bq} in the context of binary nucleation and more recently by Iwamatus~\cite{Iwamatsu:2012hy,Iwamatsu:2012bh} in the context of competitive nucleation through parallel channels, which is the case in the current model. In particular, the important challenge is to account for possible anisotropy in the rates at which monomers for the different components attach to the critical cluster that may cause the cluster to grow in a direction that differs from the steepest descent pathway through the saddle point~\cite{Reiss:1950dd}.

Here, we will simply highlight the key results from these earlier works necessary for calculating the rates and will follow the development outlined by Iwamatus~\cite{Iwamatsu:2012hy}. The Gibbs free energy for forming a cluster, $\Delta G(n_B,n_C)$ can be expanded around the saddle point associated with the critical cluster $n^*_{\kappa}=(n^*_{\kappa,B},n^*_{\kappa,C})$, where $\kappa = B,C$ denotes the identity of the saddle point leading to a particular phase, to yield~\cite{Trinkaus:1983eg},
\begin{equation}
\Delta G(n_B,n_C)\approx \Delta G^*_{\kappa}+\frac{1}{2}\sum_{i,j}(n_i-n^*_{\kappa,i})\Delta G^*_{\kappa,ij}(n_j-n^*_{\kappa,j})\mbox{,}\\
\label{eq:gexpand}
\end{equation}
where $\Delta G^*_{\kappa}=\Delta G(n^*_{\kappa})$ is the height of the free energy barrier at the saddle point for the transition to phase $\kappa$ and
\begin{equation}
\Delta G^*_{\kappa,ij}=\left(\frac{\partial^2\Delta G}{\partial n_i \partial n_j}\right)_{n^*_{\kappa}}\\
\label{eq:spc}
\end{equation}
To account for difference in the rates of adding a monomer of component $i$ to the critical cluster at the saddle point, $R^*_{\kappa,i}$, Trinkaus~\cite{Trinkaus:1983eg} and Wilemski ~\cite{Wilemski:1999bq} define the matrix element,
\begin{equation}
\Gamma^*_{\kappa,ij}=(R^*_{\kappa,i})^{1/2}\Delta G^*_{\kappa,ij}(R^*_{\kappa,j})^{1/2}\mbox{,}\\
\label{eq:rmat}
\end{equation}
which contains information about the direction of cluster growth and the curvature of the free energy surface at the saddle point. The transition state theory rate of nucleation is then expressed as
\begin{equation}
J_{\kappa}^{TST}=\sqrt{\frac{R^*_{\kappa,B}R^*_{\kappa,C}|\lambda_{\kappa}|}{\gamma_{\kappa}}}N_s\exp[-\Delta G^*_{\kappa}/kT]\mbox{,}\\
\label{eq:ratetst}
\end{equation}
where the eigenvalues $\lambda_{\kappa}$ and $\gamma_{\kappa}$ are given by
\begin{equation}
\lambda_{\kappa}=(\Gamma_{\kappa,BB}+\Gamma_{\kappa,CC}-M_{\kappa})/2 <0\mbox{,}\\
\label{eq:lk}
\end{equation}
\begin{equation}
\gamma_{\kappa}=(\Gamma_{\kappa,BB}+\Gamma_{\kappa,CC}+M_{\kappa})/2 >0\mbox{,}\\
\label{eq:gk}
\end{equation}
and
\begin{equation}
M_{\kappa}=\left((\Gamma_{\kappa,BB}-\Gamma_{\kappa,CC})^2+4(\Gamma_{\kappa,BC})^2)\right)^{1/2} \mbox{.}\\
\label{eq:m}
\end{equation}
Equation~\ref{eq:ratetst} is the binary analogue of the one dimensional TST for nucleation where $N_s\exp(-\beta\Delta G)$ is the probability of finding a cluster in the transition state and $N_s$ is the number of heterogeneous nucleation sites, which is unity in the current case.

To calculate the nucleation free energy surface, we begin by defining a conditional partition function for the system with fixed $N, h_B, h_C, T$, as $Z(n_B,n_C)=\sum \exp[-\beta E(\sigma)]$, where the sum is over all configurations that contain an $(n_B,n_C)$--cluster. The full partition function is then obtained by summing over all possible clusters, $Z=\sum_{ n_B}\sum_{n_C}Z(n_B,n_C)$ and the probability of observing a cluster is $P(n_B,n_C)=Z(n_B,n_C)/Z$. Ten Wolde {\it et} {\it al}~\cite{ReinTenWolde:1996p1389} defined the free energy barrier to nucleation as the minimum reversible work required to constrain the metastable system to the transition state. In the case of heterogeneous nucleation onto a microscopic impurity, where the cluster size represents a well defined order parameter that describes the microscopic state of the system, the partition function for the metastable state can be expressed as a sum over all cluster sizes smaller than the critical cluster~\cite{Scheifele:2013uo}. In the present case, this can be written, 
\begin{equation}
Z_m=\sum_{n_{B},n_{C}=0}^{(n_B^{\prime},n_C^{\prime})} Z(n_B,n_C) \mbox{,}\\
\label{eq:zm}
\end{equation}
where the ($n_B^{\prime},n_C^{\prime})$--clusters denote the boundary on the two dimensional free energy surface that separates the those clusters that tend to grow and those that tend to shrink. The work of forming a cluster is then,
\begin{eqnarray}
\beta\Delta G(n_B,n_C) &=-\ln \frac{Z(n_B,n_C)}{Z_m}\nonumber\\
&=-\ln \frac{P(n_B,n_C)}{\sum_{n_B,n_C}^{(n_B^{\prime},n_C^{\prime})}P(n_B,n_C)}\mbox{.}
\label{eq:dg}
\end{eqnarray}
We employ biased umbrella sampling~\cite{Frenkel2002} MC simulations to calculate $P(n_B,n_C)$, using a parabolic biasing potential, $U_0= c(n_B-n_{0B})^2+c(n_C-n_{0C})^2$, where $n_{0B}$ and $n_{0C}$ denote the umbrella center for a simulation window and $c$ is a constant that controls the strength of the bias. To access the entire free energy landscape of the metastable region, we use a $12\times12$ grid of umbrella windows where $n_{0B}$ and $n_{0C}$ range from 0 to 110 at intervals of 10. 

One Monte Carlo step (mcs) is equivalent to $N=L^2$ flip attempts, where each flip attempt is accepted with a probability $min\left\{1,exp(-\beta\Delta E)\right\}$. We evaluate the largest cluster containing the impurity after each mcs, and then apply the constrained potential. For each umbrella center, we run the simulation for $6.4 \times 10^5$ mcs, saving the cluster size ($n_B,n_C$) after each 250 mcs. During analysis we drop the initial 10000 mcs from the statistics to allow for proper equilibration. The Multiple Bernett Acceptance Ratio (MBAR) estimator \cite{Shirts:2008p1387} is used to construct the full free energy surface from the data obtained in different umbrella windows. 

The saddle point properties for each transition, such as $n^*_{\kappa}$ and $\Delta G^*_{\kappa,ij}$, are obtained by fitting the free energy in the saddle point region to a two dimensional quadratic function, $\beta\Delta G(n_B,n_C)=a(n_B-n^*_{\kappa,B})^2+b(n_C-n^*_{\kappa,C})^2+c(n_B-n^*_{\kappa,B})(n_C-n^*_{\kappa,C})+d$, where $a,b,c,d,n^*_{\kappa,B}$ and $n^*_{\kappa,C}$ are all fit parameters. To calculate the rates of attachment, we adapt the method developed by Frenkel {\it et al.}~\cite{Auer:2004gr} for a single component system and assume that the diffusion in cluster size with respect to the two components are independent of each other, which gives,
\begin{equation}
R_{\kappa,i}=\frac{\left<[n_i(t)-n^*_{\kappa,i}(0)]^2\right>}{2t}\mbox{,}\\
\label{eq:rki}
\end{equation}
and the ensemble average is taken over 10 MC simulation trajectories that start with independent critical clusters obtained from our biased umbrella sampling simulations and are terminated when $\left| n_i(t)\right| > 10$. 

\section{Results and Discussion}
\subsection{Survival Probability}
\begin{figure}[]
\includegraphics[width=3.5in]{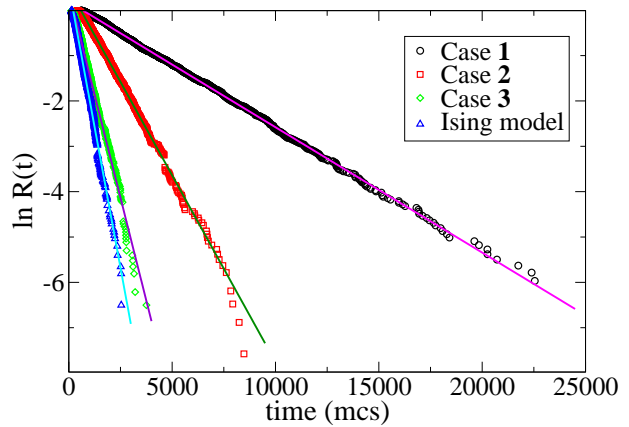}
\caption{$\ln R(t)$ as a function of time for all three competitive nucleation cases and the single component Ising model. The points represent the simulation data and the solid lines represent fits of Eq.~\ref{eq:esp}.}
\label{fig:SP}
\end{figure}

Following the cluster size, $(n_B,n_C)$, as a function of time during a trajectory shows that the system exhibits typical nucleation behaviour with the cluster's size fluctuating, growing and shrinking, until it eventually nucleates and grows to consume the system. In particular, we do not see any trajectories where the cluster size decreases again once it has reached the threshold, $n_B$ or $n_C=150$, suggesting our criteria clearly identifies a nucleation event. Beyond this requirement, the rate obtained by fitting Eq.~\ref{eq:esp} to the data (Fig.~\ref{fig:SP}) is insensitive to detailed location of the threshold~\cite{SaikaVoivod:2006p6}.

\begin{figure}[]
\includegraphics[width=3.5in]{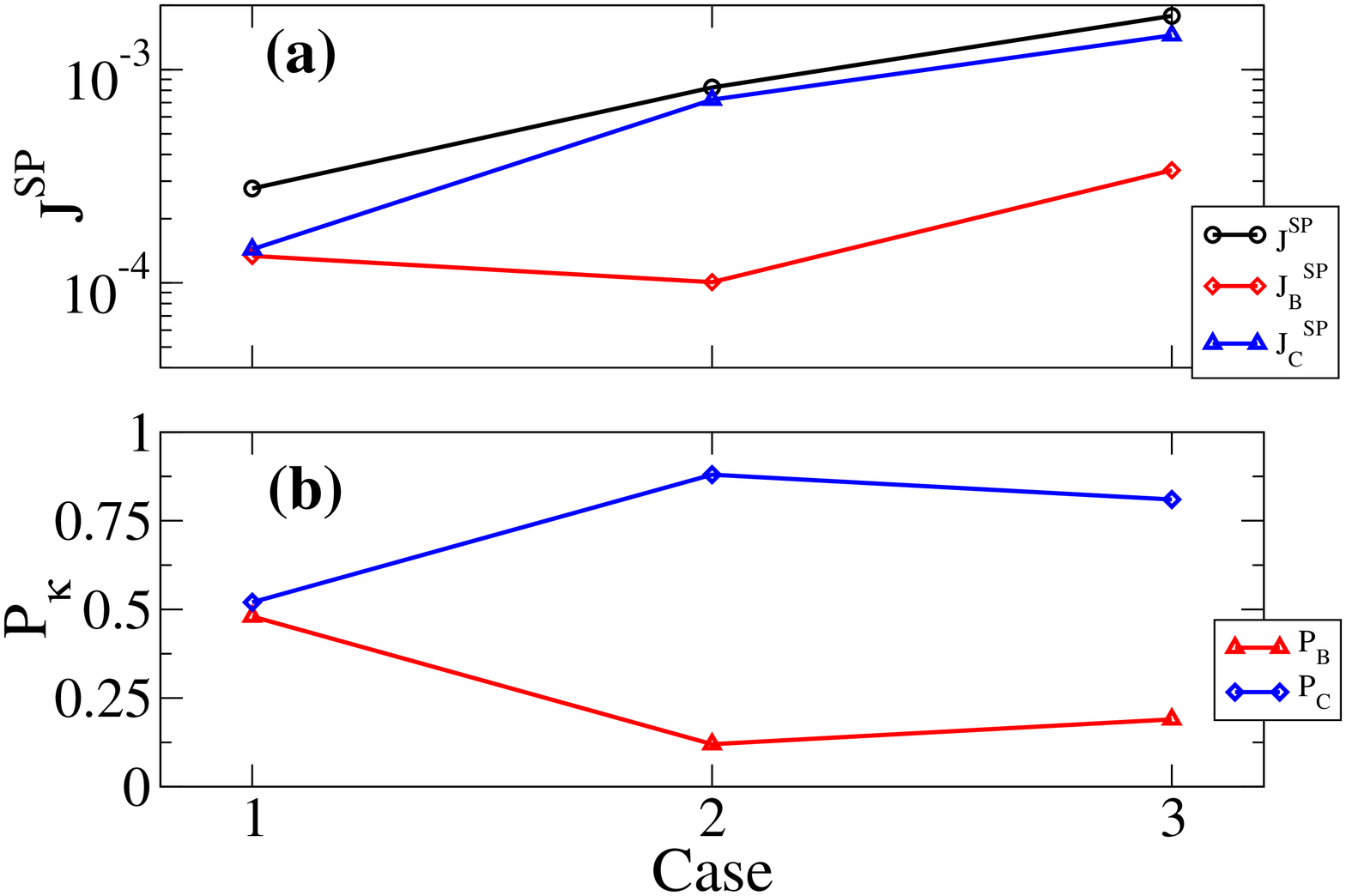}
\caption{Survival probability analysis. (a) The nucleation rates, $J^{SP}$, $J^{SP}_{B}$ and $J^{SP}_{C}$ in units of mcs$^{-1}$ for each case. (b) The nucleation probability, $P_{\kappa}$, for phases B and C for each case. }
\label{fig:sp_prob}
\end{figure}

Figure~\ref{fig:sp_prob} shows the nucleation rates and probabilities obtained from the survival probability analysis. A full set of results is presented in Table~\ref{tab:summary}. {\bf Case 1} has the lowest overall rate and the probabilities of observing nucleation for the two stable phases is the same because the thermodynamic driving force, determined by the field strength, and the surface interactions between phases, determined by the spin interactions, are equal. Increasing the field strength that favours phase $C$ ({\bf Case 2}) increases the overall rate by inducing a large increase in $J^{SP}_C$ that outweighs the small decrease observed in $J^{SP}_B$. These rate changes shift the product distribution to favour C by $88:12$. In moving from {\bf Case 2} to {\bf Case 3}, we have reduced the degree of repulsion between the two stable phases which should lower the surface tension between the phases and increase the degree of mixing. This results in an increase in all the rates, with the largest increase occurring in $J^{SP}_B$. It also causes a small change in the product distribution.

\begin{table}[h]
\caption{\label{tab:summary}Summary of simulation results.}
\begin{ruledtabular}
\begin{tabular}{llll} 
Property 	& {\bf Case 1} 	& {\bf Case 2} 	& {\bf Case 3} \\ \hline
$P_B$ (SP)			& 0.48	& 0.12	& 0.19\\
$P_C$ (SP)			& 0.52	& 0.88	& 0.81\\
$J^{SP}\times 10^{4}$	&2.8		&8.2		&17.8\\
$J^{SP}_B\times 10^{4}$	&1.3		&1.0		&3.4\\
$J^{SP}_C\times 10^{4}$	&1.4		&7.2		&14.4\\
$\beta \Delta G^*_B$			&8.9		&9.8		&8.6\\
$\beta \Delta G^*_C$			&8.8		&8.0		&7.3\\
$n^*_B$				&69.3, 4.5	&68.1, 4.7	&57.2, 9.2\\
$n^*_C$				&4.5, 72	&3.4, 47.0	&6.1, 38.5\\
$R_{B,B},R_{B,C}$		&11.9, 0.8	&19.6, 6.8 &8.9, 7.1\\
$R_{C,B},R_{C,C}$		&1.6, 13.7	&1.3, 18.5 &1.7, 10.9\\
$J^{TST}_B\times 10^{4}$&1.8		&1.2		&4.4\\
$J^{TST}_C\times 10^{4}$&1.8		&9.3		&20.8\\
$J^{TST}\times 10^{4}$	&3.6		&10.5	&25.2\\
\end{tabular}
\end{ruledtabular}
\end{table}

\subsection{Free energy Surfaces}
\begin{figure}[htp]
\subfloat[{\bf Case 1}]{%
\includegraphics[width=3.0in]{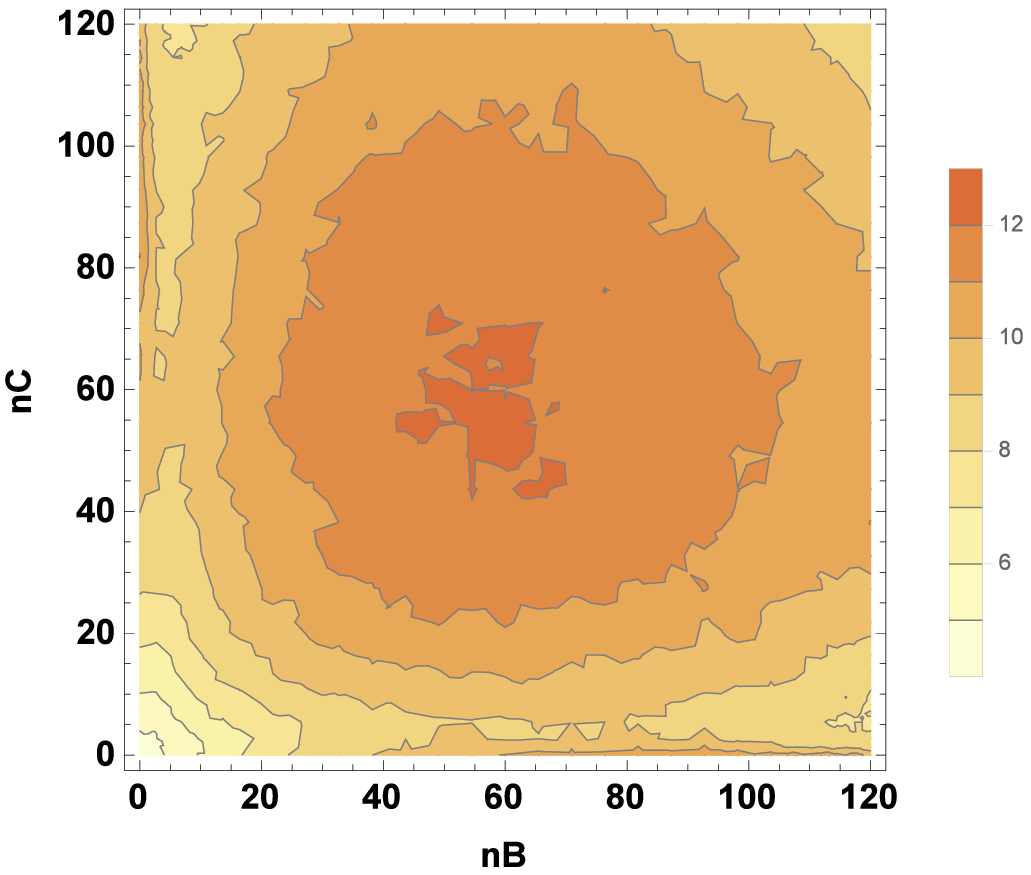}%
}

\subfloat[{\bf Case 2}]{%
\includegraphics[width=3.0in]{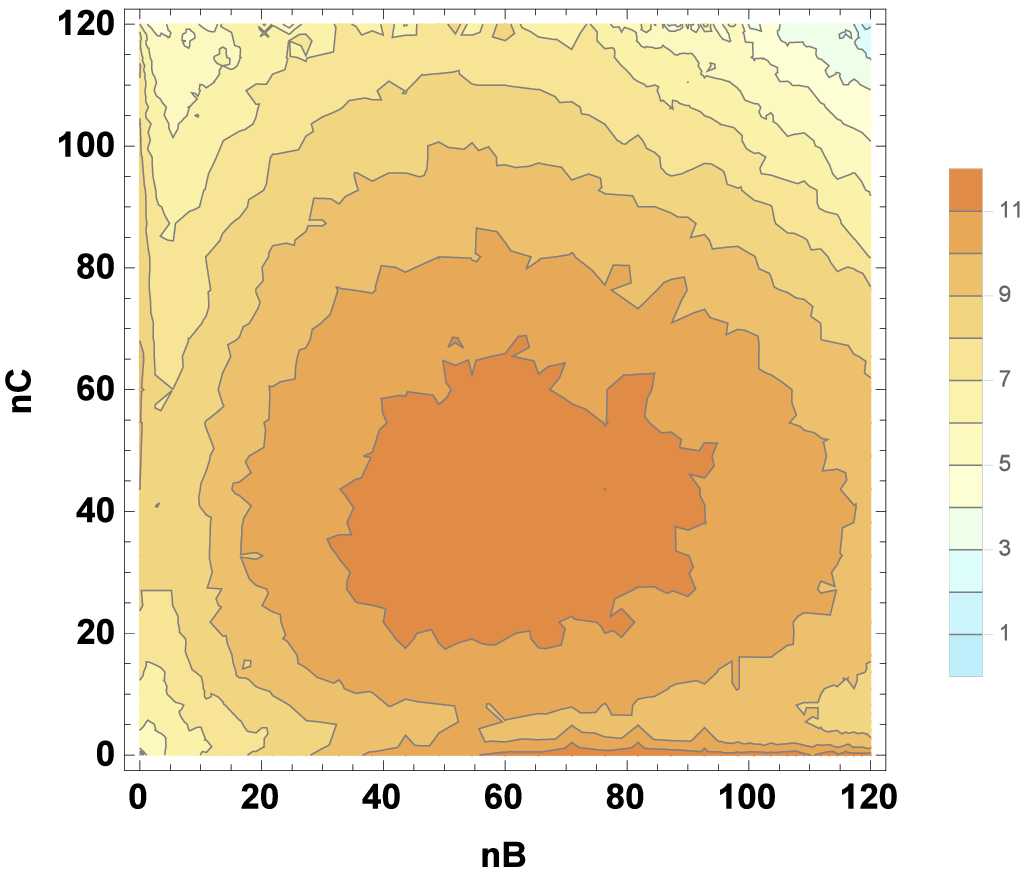}%
}

\subfloat[{\bf Case 3}]{%
\includegraphics[width=3.0in]{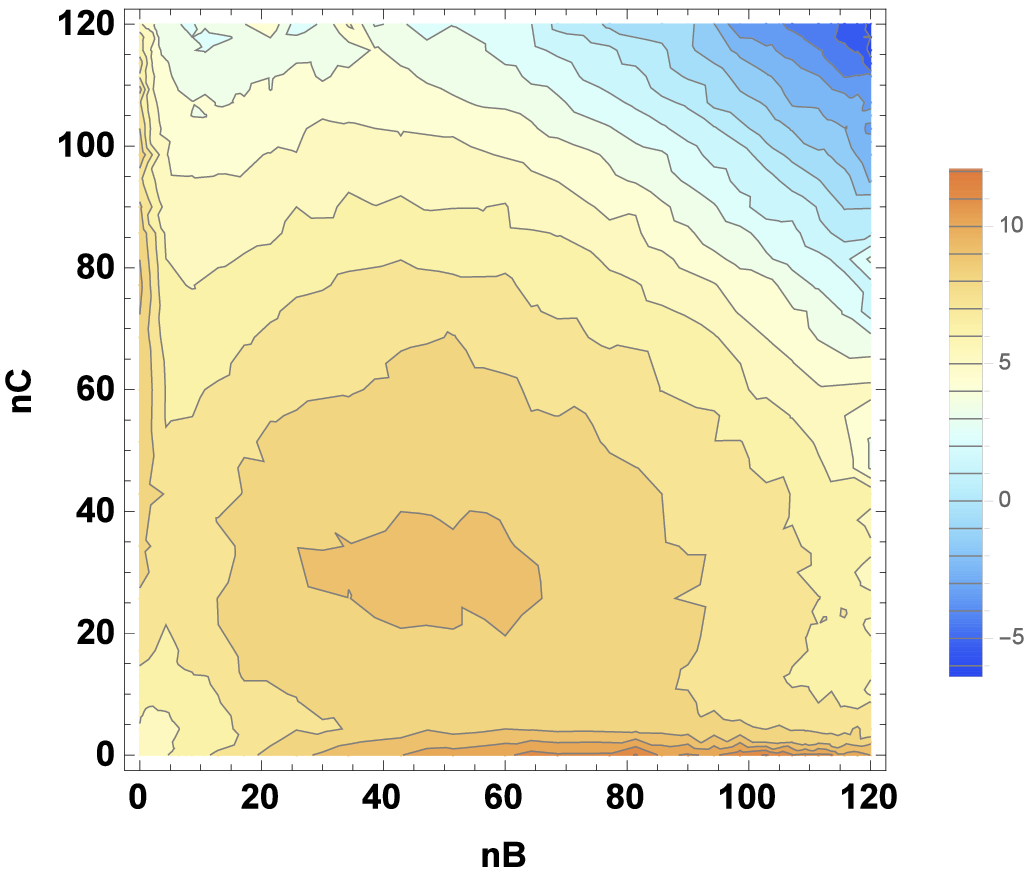}%
}
\caption{Countour plots of the $\beta\Delta G(n_B,n_C)$ free energy surface for the three cases studied.}
\label{fig:contour}
\end{figure}

Figure~\ref{fig:contour} shows the contour plots of the free energy surfaces for all three cases. For the purposes of calculating $\beta \Delta G(n_B,n_C)$, we define the metastable basin as the rectangular region bounded by the zero sized cluster as the lower limit and the largest component of the critical nuclei for each phase as the upper bounds along each axis. In principle, it is possible to define the metastable region more rigorously by identifying the appropriate ridges and valleys on the surface based on its curvature~\cite{Hoffman:1986il}. However, the boundary region away from the saddle points themselves only contributes a small amount to the Boltzmann weighted configuration space of the metastable phase and we would expect our results to be relatively insensitive to small changes in how it is defined. A key feature of the free energy normalization described in Eqs.~\ref{eq:zm} and \ref{eq:dg} is that $\beta \Delta G(n_B,n_C)$ represents the work of forming an $(n_B,n_C)$-embryo out of the metastable phase. As a result, $\beta \Delta G(0,0)\neq 0$ because it requires work to constrain the metastable state to a region of phase space containing only the bare impurity. Similarly, $\beta \Delta G(n_B,n_C) > 0$ for all clusters in the metastable region. Defined in this way, $\beta \Delta G^*_{\kappa}$ is directly related to the probability of finding the system in the transition state~\cite{Scheifele:2013uo}. 

The Gibbs free energy surfaces for all cases exhibit two nucleation channels, running parallel along the $n_B$ and $n_C$ axes respectively, that exit the metastable basin through a saddle point. As expected, the composition of the critical nuclei are dominated by the nucleating phase but there is always a small number of spins associated with the competing phase present as shown in Figs.~\ref{fig:pott}b-d. For example, the critical nucleus for the $C$ transition in {\bf Case 1}, only contains a 5\% number faction of the $B$ (See Table~\ref{tab:summary}) and this fraction grows slightly to 13\% as the repulsive interaction is reduced ({\bf Case 3}). 

The free energy barriers and critical nuclei sizes are the same for both transitions in {\bf Case 1} (e.g.  $\beta \Delta G_C^*=8.8$ and $n_C^*=[4.5,72]$), then for the $C$ transition in {\bf Case 2} we see $\beta\Delta G^*_C$ decrease by 0.8 and the critical size decreases by 25 spins, consistent with the expected increase in the nucleation rate. We also note that the barrier height, $\beta\Delta G^*_B$, increases by 0.9 relative to {\bf Case 1}, which is consistent with the decrease in the rate for the B-tansition. However, there is no change in its critical nucleus size and we have not altered the driving force with respect to phase $B$. To understand this, we note that the difference in free energy between the saddle point for the $B$-transition and the dry impurity, $\Delta G^*_B-\Delta G(0,0)$, does not change between {\bf Case 1} and {\bf Case 2}, meaning that both free energies have increased. This suggests that it is a change in the normalization that causes the barrier to increase, which highlights the fact that a change in the driving force for one phase not only influences the barrier for that transition, but it also influences the properties of the metastable state as a whole, thus influencing the properties of the competing processes as well. The barriers and critical nuclei sizes to both transitions decrease in {\bf Case 3}, but we see a greater decrease in $\Delta G^*_B$, possibly because a reduction in the repulsive interactions between phases $B$ and $C$ should decrease the surface tension which would benefit the larger clusters more.

\subsection{Effects of Competitive Nucleation}
Our simple model can now be used to explore how the competition between different phases attempting to nucleate on a microscopic impurity affect nucleation. We begin by comparing the nucleation rate obtained in {\bf Case 1}, which involves the competitive nucleation of two equally stable phases, with the rate obtained for the Ising model, where there is just a single stable phase. Both stables phases in {\bf Case 1} have identical properties to those of the stable Ising model phase, with the same interactions with the metastable phase and the impurity. The phases also all share the same thermodynamic driving force. However, Fig.~\ref{fig:SP} shows that the survival probability for the Ising model decays much more rapidly than the competitive nucleation case with a measured nucleation rate that is nine times faster.

\begin{figure}[]
\includegraphics[width=3.5in]{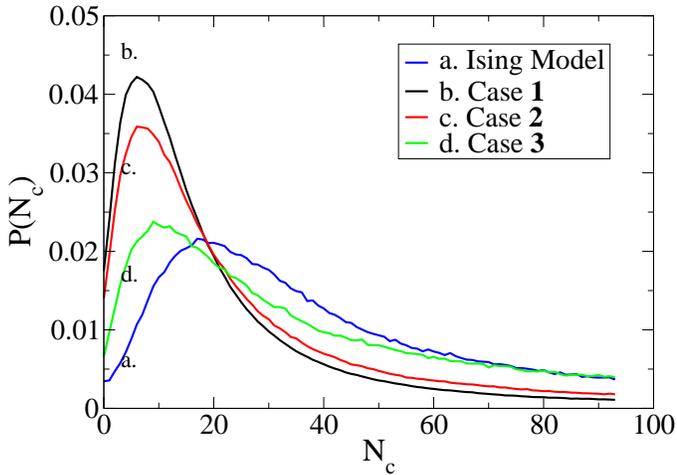}
\caption{The probability, $P(N_c)$ of observing an $N_c=n_B+n_C$ sized embryo on the impurity for all three competitive nucleation cases studied and the Ising model.}
\label{fig:wetting}
\end{figure}

Scheifele $et$ $al$~\cite{Scheifele:2013uo} calculated the nucleation free energy surface for heterogeneous nucleation in the Ising model under similar conditions to those studied here. They found the free energy exhibited a local minimum prior to the critical embryo size that shows the system spontaneously forms a wetting layer surrounding the impurity, before it eventually nucleated to grow the droplet. This feature is absent from our free energy surfaces (Fig.~\ref{fig:contour}) and the bare impurity, with $n_B,n_C=0$, is the lowest free energy state in the metastable phase. To examine the possibility of a wetting layer more closely, we also calculate the probability, $P(N_c)$,  of finding an $N_c=n_B+n_C$ cluster on the impurity for all the cases studied and the Ising model. Figure~\ref{fig:wetting} shows that $P(N_c)$ has a maximum at around $N_c\approx20$ for the Ising model, which corresponds to a single layer completely wetting the linear seven spin impurity. The competitive nucleation case exhibits a narrow distribution around a peak maximum at $N_c\approx 9$ which represents a sub-monolayer but when we reduce the repulsion, as in {\bf Case 3}, we see the peak broaden and begin to shift to larger $N_c$, indicating an increase in the wetting. This clearly shows that competition between two phases attempting to nucleate on an impurity slows the rate relative to the nucleation of a single phase and that competition leads to an effective drying of the impurity because the repulsive interaction between the phases introduces a high energy cost for the phases to coexist. 

In addition, the volume of configuration space associated with the metastable phases increases in the competitive nucleation case because of the multicomponent nature of the clusters and this would be expected to contribute to the slowing of the nucleation rate relative to the single component case.
Sear~\cite{Sear:2005ca} also used a similar Potts model, along with a simple CNT model, to show that the properties of the impurity, through its interaction with the different phases, could influence the order in which the phases nucleated. However, the interaction between the phases was not investigated and we would expect this aspect to become important when the two competing phases have high surface free energies, which would be the case of competitive freezing of two different crystals.

\subsection{Comparing Methods}
\begin{figure}[]
\includegraphics[width=3.5in]{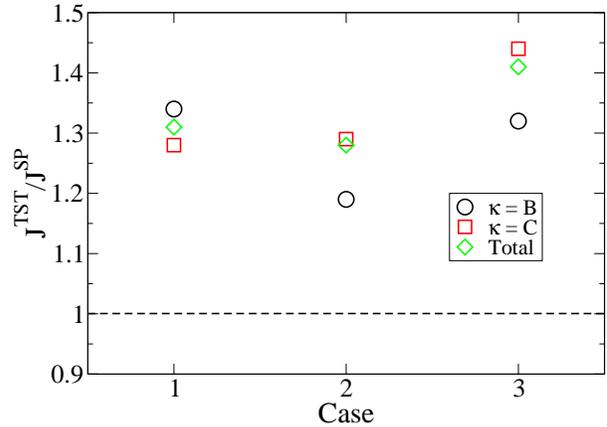}
\caption{The TST nucleation rate relative to the SP nucleation rate for the transitions to phases $\kappa = B$ and $C$, and the total rate.}
\label{fig:compare}
\end{figure}

The absolute rates obtained from our TST calculations follow the same trends as those obtained through the survival probabilities, so it is more useful to compare the relative rates of the two independent calculations. Figure~\ref{fig:compare} shows that $J^{TST}$ is within 50\% of $J^{SP}$ for all the cases studied, but they are also consistently higher. An important feature of the survival probability approach is that our criteria for identifying a nucleation event is set well beyond the critical boundary, which includes the saddle. Trajectories that initially cross critical boundary, but then cross back into the metastable region before reaching the nucleation criteria are not considered to have reacted until they eventually cross the nucleation criteria at a later time. This recrossing leads to a longer average nucleation time and a slower rate. In contrast, the TST approach assumes there are no recrossing events, so that an embryo is considered to have nucleated once it crosses the critical boundary. As a result, $J^{TST}>J^{SP}$ as observed.

\begin{figure}[]
\includegraphics[width=3.5in]{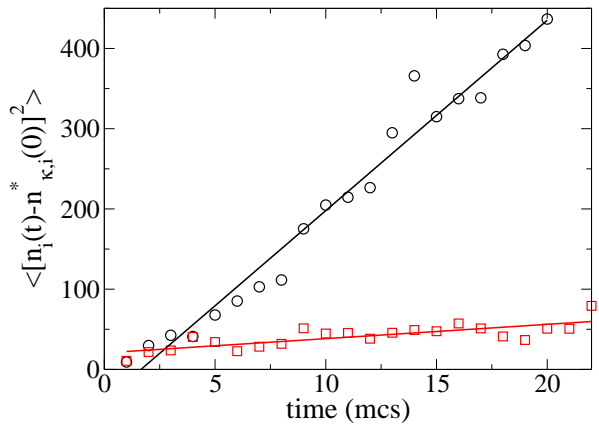}
\caption{The mean squared nuclei composition for component $B$ (circles) and $C$ (squares) as a function of time for the $\kappa=B$ saddle point in {\bf Case 1}. The solid lines represent linear fits to the data. }
\label{fig:attach}
\end{figure}

The calculation of the monomer attachment rates, $R_{\kappa,i}$, also play an important role in obtaining the TST predictions. These are given by Eq.~\ref{eq:rki} and can be effectively obtained by considering the mean squared displacement in the cluster composition at a saddle point and dividing the slope of the best fit line for a given component by two (see Fig.~\ref{fig:attach}). Not surprisingly, the largest monomer attachment rates generally occur in the direction parallel to the phase being nucleated. The monomer attachment for the competing phases, which grow orthogonally, are much smaller and their small slopes suggest there may be a greater degree of error in our estimation of these quantities. The $R_{\kappa,i}$ for all cases and saddle points are reported in Table~\ref{tab:summary}.

\section{Conclusions}
In this work, we have shown that competition between two phases attempting to nucleate on an impurity influences nucleation by slowing down the nucleation rate and causing an effective drying of the substrate when the two stable phases have a repulsive interaction energy, which has potential implications for the heterogeneous nucleation of materials that exhibit polymorphism, such as the nucleation of water in the atmosphere. We also show that transition state theory for nucleation, which involves the calculation of the free energy barrier to nucleation, along with the monomer attachment rates for the two components, predicts the correct nucleation rate when the free energy has been correctly normalized with respect to the metastable state.

\acknowledgments
We would like to thank P. H. Poole and I. Sakia-Voivod for helpful discussions. We would also like to thank Compute Canada and WestGrid for providing computational resources and 
the Natural Sciences and Engineering Research Council of Canada (NSERC) for financial support.


%



\begin{thebibliography}{40}%
\makeatletter
\providecommand \@ifxundefined [1]{%
 \@ifx{#1\undefined}
}%
\providecommand \@ifnum [1]{%
 \ifnum #1\expandafter \@firstoftwo
 \else \expandafter \@secondoftwo
 \fi
}%
\providecommand \@ifx [1]{%
 \ifx #1\expandafter \@firstoftwo
 \else \expandafter \@secondoftwo
 \fi
}%
\providecommand \natexlab [1]{#1}%
\providecommand \enquote  [1]{``#1''}%
\providecommand \bibnamefont  [1]{#1}%
\providecommand \bibfnamefont [1]{#1}%
\providecommand \citenamefont [1]{#1}%
\providecommand \href@noop [0]{\@secondoftwo}%
\providecommand \href [0]{\begingroup \@sanitize@url \@href}%
\providecommand \@href[1]{\@@startlink{#1}\@@href}%
\providecommand \@@href[1]{\endgroup#1\@@endlink}%
\providecommand \@sanitize@url [0]{\catcode `\\12\catcode `\$12\catcode
  `\&12\catcode `\#12\catcode `\^12\catcode `\_12\catcode `\%12\relax}%
\providecommand \@@startlink[1]{}%
\providecommand \@@endlink[0]{}%
\providecommand \url  [0]{\begingroup\@sanitize@url \@url }%
\providecommand \@url [1]{\endgroup\@href {#1}{\urlprefix }}%
\providecommand \urlprefix  [0]{URL }%
\providecommand \Eprint [0]{\href }%
\providecommand \doibase [0]{http://dx.doi.org/}%
\providecommand \selectlanguage [0]{\@gobble}%
\providecommand \bibinfo  [0]{\@secondoftwo}%
\providecommand \bibfield  [0]{\@secondoftwo}%
\providecommand \translation [1]{[#1]}%
\providecommand \BibitemOpen [0]{}%
\providecommand \bibitemStop [0]{}%
\providecommand \bibitemNoStop [0]{.\EOS\space}%
\providecommand \EOS [0]{\spacefactor3000\relax}%
\providecommand \BibitemShut  [1]{\csname bibitem#1\endcsname}%
\let\auto@bib@innerbib\@empty
\bibitem [{\citenamefont {Debenedetti}(1996)}]{Debenedetti:1996tf}%
  \BibitemOpen
  \bibfield  {author} {\bibinfo {author} {\bibfnamefont {P.~G.}\ \bibnamefont
  {Debenedetti}},\ }\href
  {http://books.google.com/books?hl=en&lr=&id=tzvvsltE6Y8C&oi=fnd&pg=PR11&dq=Debenedetti+PG+(Book+Metastable+liquids)&ots=d7bcbhBgK&sig=F9jzABGPKBUSkjkywvG4QctRgmc}
  {\emph {\bibinfo {title} {Metastable Liquids: Concepts and Principles.}}}\
  (\bibinfo  {publisher} {Princeton University Press},\ \bibinfo {year}
  {1996})\BibitemShut {NoStop}%
\bibitem [{\citenamefont {Kelton}\ and\ \citenamefont
  {Greer}(2010)}]{Kelton:2010wj}%
  \BibitemOpen
  \bibfield  {author} {\bibinfo {author} {\bibfnamefont {K.~F.}\ \bibnamefont
  {Kelton}}\ and\ \bibinfo {author} {\bibfnamefont {A.~L.}\ \bibnamefont
  {Greer}},\ }\href {http://www.worldcat.org/title/nucleation/oclc/474342766}
  {\emph {\bibinfo {title} {{Nucleation in Condensed Matter: Applications in
  Materials and Biology}}}},\ \bibinfo {series} {Pergamon Materials Series},
  Vol.~\bibinfo {volume} {15}\ (\bibinfo  {publisher} {Elsevier},\ \bibinfo
  {address} {Oxford, New York},\ \bibinfo {year} {2010})\BibitemShut {NoStop}%
\bibitem [{\citenamefont {Chushak}\ and\ \citenamefont
  {Bartell}(2001)}]{Chushak:2001p1372}%
  \BibitemOpen
  \bibfield  {author} {\bibinfo {author} {\bibfnamefont {Y.~G.}\ \bibnamefont
  {Chushak}}\ and\ \bibinfo {author} {\bibfnamefont {L.~S.}\ \bibnamefont
  {Bartell}},\ }\href {http://pubs.acs.org/doi/abs/10.1021/jp0109426}
  {\bibfield  {journal} {\bibinfo  {journal} {J. Phys. Chem. B}\ }\textbf
  {\bibinfo {volume} {105}},\ \bibinfo {pages} {11605} (\bibinfo {year}
  {2001})}\BibitemShut {NoStop}%
\bibitem [{\citenamefont {Asuquo}\ and\ \citenamefont
  {Bowles}(2012)}]{Asuquo:2012p1373}%
  \BibitemOpen
  \bibfield  {author} {\bibinfo {author} {\bibfnamefont {C.~C.}\ \bibnamefont
  {Asuquo}}\ and\ \bibinfo {author} {\bibfnamefont {R.~K.}\ \bibnamefont
  {Bowles}},\ }\href {http://pubs.acs.org/doi/abs/10.1021/jp2115274} {\bibfield
   {journal} {\bibinfo  {journal} {J. Phys. Chem. C}\ }\textbf {\bibinfo
  {volume} {116}},\ \bibinfo {pages} {14619} (\bibinfo {year}
  {2012})}\BibitemShut {NoStop}%
\bibitem [{\citenamefont {Koga}\ \emph {et~al.}(2004)\citenamefont {Koga},
  \citenamefont {Ikeshoji},\ and\ \citenamefont {Sugawara}}]{Koga:2004cs}%
  \BibitemOpen
  \bibfield  {author} {\bibinfo {author} {\bibfnamefont {K.}~\bibnamefont
  {Koga}}, \bibinfo {author} {\bibfnamefont {T.}~\bibnamefont {Ikeshoji}}, \
  and\ \bibinfo {author} {\bibfnamefont {K.}~\bibnamefont {Sugawara}},\ }\href
  {\doibase 10.1103/PhysRevLett.92.115507} {\bibfield  {journal} {\bibinfo
  {journal} {Phys. Rev. Lett.}\ }\textbf {\bibinfo {volume} {92}},\ \bibinfo
  {pages} {115507} (\bibinfo {year} {2004})}\BibitemShut {NoStop}%
\bibitem [{\citenamefont {Koga}(2006)}]{Koga:2006p12920}%
  \BibitemOpen
  \bibfield  {author} {\bibinfo {author} {\bibfnamefont {K.}~\bibnamefont
  {Koga}},\ }\href {\doibase 10.1103/PhysRevLett.96.115501} {\bibfield
  {journal} {\bibinfo  {journal} {Phys. Rev. Lett.}\ }\textbf {\bibinfo
  {volume} {96}},\ \bibinfo {pages} {115501} (\bibinfo {year}
  {2006})}\BibitemShut {NoStop}%
\bibitem [{\citenamefont {Eddaoudi}\ \emph {et~al.}(2002)\citenamefont
  {Eddaoudi}, \citenamefont {Kim}, \citenamefont {Rosi}, \citenamefont {Vodak},
  \citenamefont {Wachter}, \citenamefont {O'Keeffe},\ and\ \citenamefont
  {Yaghi}}]{Eddaoudi:2002p1673}%
  \BibitemOpen
  \bibfield  {author} {\bibinfo {author} {\bibfnamefont {M.}~\bibnamefont
  {Eddaoudi}}, \bibinfo {author} {\bibfnamefont {J.}~\bibnamefont {Kim}},
  \bibinfo {author} {\bibfnamefont {N.}~\bibnamefont {Rosi}}, \bibinfo {author}
  {\bibfnamefont {D.}~\bibnamefont {Vodak}}, \bibinfo {author} {\bibfnamefont
  {J.}~\bibnamefont {Wachter}}, \bibinfo {author} {\bibfnamefont
  {M.}~\bibnamefont {O'Keeffe}}, \ and\ \bibinfo {author} {\bibfnamefont
  {O.~M.}\ \bibnamefont {Yaghi}},\ }\href {\doibase 10.1126/science.1067208}
  {\bibfield  {journal} {\bibinfo  {journal} {Science}\ }\textbf {\bibinfo
  {volume} {295}},\ \bibinfo {pages} {469} (\bibinfo {year}
  {2002})}\BibitemShut {NoStop}%
\bibitem [{\citenamefont {Morris}\ \emph {et~al.}(2001)\citenamefont {Morris},
  \citenamefont {Griesser}, \citenamefont {Eckhardt},\ and\ \citenamefont
  {Stowell}}]{Morris:2001p1672}%
  \BibitemOpen
  \bibfield  {author} {\bibinfo {author} {\bibfnamefont {K.~R.}\ \bibnamefont
  {Morris}}, \bibinfo {author} {\bibfnamefont {U.~J.}\ \bibnamefont
  {Griesser}}, \bibinfo {author} {\bibfnamefont {C.~J.}\ \bibnamefont
  {Eckhardt}}, \ and\ \bibinfo {author} {\bibfnamefont {J.~G.}\ \bibnamefont
  {Stowell}},\ }\href@noop {} {\bibfield  {journal} {\bibinfo  {journal} {Adv.
  Drug Deliv. Rev.}\ }\textbf {\bibinfo {volume} {48}},\ \bibinfo {pages} {91}
  (\bibinfo {year} {2001})}\BibitemShut {NoStop}%
\bibitem [{\citenamefont {Murray}\ \emph {et~al.}(2005)\citenamefont {Murray},
  \citenamefont {Knopf},\ and\ \citenamefont {Bertram}}]{Murray:2005fq}%
  \BibitemOpen
  \bibfield  {author} {\bibinfo {author} {\bibfnamefont {B.~J.}\ \bibnamefont
  {Murray}}, \bibinfo {author} {\bibfnamefont {D.~A.}\ \bibnamefont {Knopf}}, \
  and\ \bibinfo {author} {\bibfnamefont {A.~K.}\ \bibnamefont {Bertram}},\
  }\href {\doibase 10.1038/nature03403} {\bibfield  {journal} {\bibinfo
  {journal} {Nature}\ }\textbf {\bibinfo {volume} {434}},\ \bibinfo {pages}
  {202} (\bibinfo {year} {2005})}\BibitemShut {NoStop}%
\bibitem [{\citenamefont {Murray}\ and\ \citenamefont
  {Bertram}(2006)}]{Murray:2006fn}%
  \BibitemOpen
  \bibfield  {author} {\bibinfo {author} {\bibfnamefont {B.~J.}\ \bibnamefont
  {Murray}}\ and\ \bibinfo {author} {\bibfnamefont {A.~K.}\ \bibnamefont
  {Bertram}},\ }\href {\doibase 10.1039/B513480C} {\bibfield  {journal}
  {\bibinfo  {journal} {Phys. Chem. Chem. Phys.}\ }\textbf {\bibinfo {volume}
  {8}},\ \bibinfo {pages} {186} (\bibinfo {year} {2006})}\BibitemShut {NoStop}%
\bibitem [{\citenamefont {Ostwald}(1897)}]{Ostwald:1897wd}%
  \BibitemOpen
  \bibfield  {author} {\bibinfo {author} {\bibfnamefont {W.}~\bibnamefont
  {Ostwald}},\ }\href@noop {} {\bibfield  {journal} {\bibinfo  {journal} {Z.
  Phys. Chem. A}\ }\textbf {\bibinfo {volume} {22}},\ \bibinfo {pages} {289}
  (\bibinfo {year} {1897})}\BibitemShut {NoStop}%
\bibitem [{\citenamefont {Stranski}\ and\ \citenamefont
  {Totomanow}(1933)}]{stranksi1933}%
  \BibitemOpen
  \bibfield  {author} {\bibinfo {author} {\bibfnamefont {I.~N.}\ \bibnamefont
  {Stranski}}\ and\ \bibinfo {author} {\bibfnamefont {D.}~\bibnamefont
  {Totomanow}},\ }\href@noop {} {\bibfield  {journal} {\bibinfo  {journal} {Z.
  Phys. Chem. A}\ }\textbf {\bibinfo {volume} {163}},\ \bibinfo {pages} {399}
  (\bibinfo {year} {1933})}\BibitemShut {NoStop}%
\bibitem [{\citenamefont {van Meel}\ \emph {et~al.}(2008)\citenamefont {van
  Meel}, \citenamefont {Page}, \citenamefont {Sear},\ and\ \citenamefont
  {Frenkel}}]{vanMeel:2008p8395}%
  \BibitemOpen
  \bibfield  {author} {\bibinfo {author} {\bibfnamefont {J.~A.}\ \bibnamefont
  {van Meel}}, \bibinfo {author} {\bibfnamefont {A.~J.}\ \bibnamefont {Page}},
  \bibinfo {author} {\bibfnamefont {R.~P.}\ \bibnamefont {Sear}}, \ and\
  \bibinfo {author} {\bibfnamefont {D.}~\bibnamefont {Frenkel}},\ }\href
  {\doibase 10.1063/1.3026364} {\bibfield  {journal} {\bibinfo  {journal} {J.
  Chem. Phys.}\ }\textbf {\bibinfo {volume} {129}},\ \bibinfo {pages} {204505}
  (\bibinfo {year} {2008})}\BibitemShut {NoStop}%
\bibitem [{\citenamefont {Chung}\ \emph {et~al.}(2009)\citenamefont {Chung},
  \citenamefont {Kim}, \citenamefont {Kim},\ and\ \citenamefont
  {Kim}}]{Chung:2009fe}%
  \BibitemOpen
  \bibfield  {author} {\bibinfo {author} {\bibfnamefont {S.~Y.}\ \bibnamefont
  {Chung}}, \bibinfo {author} {\bibfnamefont {Y.~M.}\ \bibnamefont {Kim}},
  \bibinfo {author} {\bibfnamefont {J.~G.}\ \bibnamefont {Kim}}, \ and\
  \bibinfo {author} {\bibfnamefont {Y.~J.}\ \bibnamefont {Kim}},\ }\href
  {\doibase 10.1038/nphys1148} {\bibfield  {journal} {\bibinfo  {journal}
  {Nature Phys.}\ }\textbf {\bibinfo {volume} {5}},\ \bibinfo {pages} {68}
  (\bibinfo {year} {2009})}\BibitemShut {NoStop}%
\bibitem [{\citenamefont {Peng}\ \emph {et~al.}(2015)\citenamefont {Peng},
  \citenamefont {Wang}, \citenamefont {Wang}, \citenamefont {Alsayed},
  \citenamefont {Zhang}, \citenamefont {Yodh},\ and\ \citenamefont
  {Han}}]{Peng:2014is}%
  \BibitemOpen
  \bibfield  {author} {\bibinfo {author} {\bibfnamefont {Y.}~\bibnamefont
  {Peng}}, \bibinfo {author} {\bibfnamefont {F.}~\bibnamefont {Wang}}, \bibinfo
  {author} {\bibfnamefont {Z.}~\bibnamefont {Wang}}, \bibinfo {author}
  {\bibfnamefont {A.~M.}\ \bibnamefont {Alsayed}}, \bibinfo {author}
  {\bibfnamefont {Z.}~\bibnamefont {Zhang}}, \bibinfo {author} {\bibfnamefont
  {A.~G.}\ \bibnamefont {Yodh}}, \ and\ \bibinfo {author} {\bibfnamefont
  {Y.}~\bibnamefont {Han}},\ }\href {\doibase 10.1038/nmat4083} {\bibfield
  {journal} {\bibinfo  {journal} {Nature Mater.}\ }\textbf {\bibinfo {volume}
  {14}},\ \bibinfo {pages} {101} (\bibinfo {year} {2015})}\BibitemShut
  {NoStop}%
\bibitem [{\citenamefont {Sear}(2009)}]{Sear:2009it}%
  \BibitemOpen
  \bibfield  {author} {\bibinfo {author} {\bibfnamefont {R.~P.}\ \bibnamefont
  {Sear}},\ }\href {\doibase 10.1063/1.3205030} {\bibfield  {journal} {\bibinfo
   {journal} {J. Chem. Phys.}\ }\textbf {\bibinfo {volume} {131}},\ \bibinfo
  {pages} {074702} (\bibinfo {year} {2009})}\BibitemShut {NoStop}%
\bibitem [{\citenamefont {Hedges}\ and\ \citenamefont
  {Whitelam}(2011)}]{Hedges:2011ci}%
  \BibitemOpen
  \bibfield  {author} {\bibinfo {author} {\bibfnamefont {L.~O.}\ \bibnamefont
  {Hedges}}\ and\ \bibinfo {author} {\bibfnamefont {S.}~\bibnamefont
  {Whitelam}},\ }\href {\doibase 10.1063/1.3655358} {\bibfield  {journal}
  {\bibinfo  {journal} {J. Chem. Phys.}\ }\textbf {\bibinfo {volume} {135}},\
  \bibinfo {pages} {164902} (\bibinfo {year} {2011})}\BibitemShut {NoStop}%
\bibitem [{\citenamefont {ten Wolde}\ and\ \citenamefont
  {Frenkel}(1997)}]{Wolde:1997ez}%
  \BibitemOpen
  \bibfield  {author} {\bibinfo {author} {\bibfnamefont {P.~R.}\ \bibnamefont
  {ten Wolde}}\ and\ \bibinfo {author} {\bibfnamefont {D.}~\bibnamefont
  {Frenkel}},\ }\href {\doibase 10.1126/science.277.5334.1975} {\bibfield
  {journal} {\bibinfo  {journal} {Science}\ }\textbf {\bibinfo {volume}
  {277}},\ \bibinfo {pages} {1975} (\bibinfo {year} {1997})}\BibitemShut
  {NoStop}%
\bibitem [{\citenamefont {Qi}\ \emph {et~al.}(2015)\citenamefont {Qi},
  \citenamefont {Peng}, \citenamefont {Han}, \citenamefont {Bowles},\ and\
  \citenamefont {Dijkstra}}]{Qi:2015ie}%
  \BibitemOpen
  \bibfield  {author} {\bibinfo {author} {\bibfnamefont {W.}~\bibnamefont
  {Qi}}, \bibinfo {author} {\bibfnamefont {Y.}~\bibnamefont {Peng}}, \bibinfo
  {author} {\bibfnamefont {Y.}~\bibnamefont {Han}}, \bibinfo {author}
  {\bibfnamefont {R.~K.}\ \bibnamefont {Bowles}}, \ and\ \bibinfo {author}
  {\bibfnamefont {M.}~\bibnamefont {Dijkstra}},\ }\href {\doibase
  10.1103/PhysRevLett.115.185701} {\bibfield  {journal} {\bibinfo  {journal}
  {Phys. Rev. Lett.}\ }\textbf {\bibinfo {volume} {115}},\ \bibinfo {pages}
  {185701} (\bibinfo {year} {2015})}\BibitemShut {NoStop}%
\bibitem [{\citenamefont {Volmer}\ and\ \citenamefont
  {Weber}(1926)}]{Volmer:1926p1353}%
  \BibitemOpen
  \bibfield  {author} {\bibinfo {author} {\bibfnamefont {M.}~\bibnamefont
  {Volmer}}\ and\ \bibinfo {author} {\bibfnamefont {A.}~\bibnamefont {Weber}},\
  }\href {\doibase 10.1103/RevModPhys.54.235} {\bibfield  {journal} {\bibinfo
  {journal} {Z. Phys. Chem.}\ }\textbf {\bibinfo {volume} {119}},\ \bibinfo
  {pages} {277} (\bibinfo {year} {1926})}\BibitemShut {NoStop}%
\bibitem [{\citenamefont {Becker}\ and\ \citenamefont
  {D{\"o}ring}(1935)}]{Becker1935pg719}%
  \BibitemOpen
  \bibfield  {author} {\bibinfo {author} {\bibfnamefont {R.}~\bibnamefont
  {Becker}}\ and\ \bibinfo {author} {\bibfnamefont {W.}~\bibnamefont
  {D{\"o}ring}},\ }\href {\doibase 10.1103/RevModPhys.54.235} {\bibfield
  {journal} {\bibinfo  {journal} {Ann. Phys.}\ }\textbf {\bibinfo {volume}
  {24}},\ \bibinfo {pages} {719} (\bibinfo {year} {1935})}\BibitemShut
  {NoStop}%
\bibitem [{\citenamefont {Zeldovich}(1943)}]{zeldovich1943}%
  \BibitemOpen
  \bibfield  {author} {\bibinfo {author} {\bibfnamefont {J.~B.}\ \bibnamefont
  {Zeldovich}},\ }\href {\doibase 10.1103/RevModPhys.54.235} {\bibfield
  {journal} {\bibinfo  {journal} {Acta Physiochimica U.S.S.R.}\ }\textbf
  {\bibinfo {volume} {18}},\ \bibinfo {pages} {1} (\bibinfo {year}
  {1943})}\BibitemShut {NoStop}%
\bibitem [{\citenamefont {Frenkel}(1959)}]{Frenkel:1959p1392}%
  \BibitemOpen
  \bibfield  {author} {\bibinfo {author} {\bibfnamefont {J.~I.}\ \bibnamefont
  {Frenkel}},\ }\href
  {http://books.google.com/books?hl=en&lr=&id=tzvvsltE6Y8C&oi=fnd&pg=PR11&dq=Debenedetti+P.+G&ots=d7fedl1GbN&sig=av0H5y9NYMpJ552iZwxR_ww-gsY}
  {\emph {\bibinfo {title} {Kinetic Theory of Liquids}}}\ (\bibinfo
  {publisher} {Dover, New York},\ \bibinfo {year} {1959})\BibitemShut {NoStop}%
\bibitem [{\citenamefont {Sanders}\ \emph {et~al.}(2007)\citenamefont
  {Sanders}, \citenamefont {Larralde},\ and\ \citenamefont
  {Leyvraz}}]{Sanders:2007p994}%
  \BibitemOpen
  \bibfield  {author} {\bibinfo {author} {\bibfnamefont {D.~P.}\ \bibnamefont
  {Sanders}}, \bibinfo {author} {\bibfnamefont {H.}~\bibnamefont {Larralde}}, \
  and\ \bibinfo {author} {\bibfnamefont {F.}~\bibnamefont {Leyvraz}},\ }\href
  {\doibase 10.1103/PhysRevB.75.132101} {\bibfield  {journal} {\bibinfo
  {journal} {Phys. Rev. B}\ }\textbf {\bibinfo {volume} {75}},\ \bibinfo
  {pages} {132101} (\bibinfo {year} {2007})}\BibitemShut {NoStop}%
\bibitem [{\citenamefont {Wu}(1982)}]{Wu:1982p1353}%
  \BibitemOpen
  \bibfield  {author} {\bibinfo {author} {\bibfnamefont {F.~Y.}\ \bibnamefont
  {Wu}},\ }\href {\doibase 10.1103/RevModPhys.54.235} {\bibfield  {journal}
  {\bibinfo  {journal} {Rev. Mod. Phys.}\ }\textbf {\bibinfo {volume} {54}},\
  \bibinfo {pages} {235} (\bibinfo {year} {1982})}\BibitemShut {NoStop}%
\bibitem [{\citenamefont {Arkin}\ and\ \citenamefont
  {Celik}(2000)}]{Arkin:2000p1125}%
  \BibitemOpen
  \bibfield  {author} {\bibinfo {author} {\bibfnamefont {H.}~\bibnamefont
  {Arkin}}\ and\ \bibinfo {author} {\bibfnamefont {T.}~\bibnamefont {Celik}},\
  }\href {http://www.worldscientific.com/doi/abs/10.1142/S0129183100001152}
  {\bibfield  {journal} {\bibinfo  {journal} {Inter. J. Mod. Phys. C}\ }\textbf
  {\bibinfo {volume} {11}},\ \bibinfo {pages} {1313} (\bibinfo {year}
  {2000})}\BibitemShut {NoStop}%
\bibitem [{\citenamefont {Scheifele}\ \emph {et~al.}(2013)\citenamefont
  {Scheifele}, \citenamefont {Saika-Voivod}, \citenamefont {Bowles},\ and\
  \citenamefont {Poole}}]{Scheifele:2013uo}%
  \BibitemOpen
  \bibfield  {author} {\bibinfo {author} {\bibfnamefont {B.}~\bibnamefont
  {Scheifele}}, \bibinfo {author} {\bibfnamefont {I.}~\bibnamefont
  {Saika-Voivod}}, \bibinfo {author} {\bibfnamefont {R.~K.}\ \bibnamefont
  {Bowles}}, \ and\ \bibinfo {author} {\bibfnamefont {P.~H.}\ \bibnamefont
  {Poole}},\ }\href {\doibase 10.1103/PhysRevE.87.042407} {\bibfield  {journal}
  {\bibinfo  {journal} {Phys. Rev. E}\ }\textbf {\bibinfo {volume} {87}},\
  \bibinfo {pages} {042407} (\bibinfo {year} {2013})}\BibitemShut {NoStop}%
\bibitem [{\citenamefont {Lundrigan}\ and\ \citenamefont
  {Saika-Voivod}(2009)}]{Lundrigan:2009p8124}%
  \BibitemOpen
  \bibfield  {author} {\bibinfo {author} {\bibfnamefont {S.~E.~M.}\
  \bibnamefont {Lundrigan}}\ and\ \bibinfo {author} {\bibfnamefont
  {I.}~\bibnamefont {Saika-Voivod}},\ }\href {\doibase 10.1063/1.3216867}
  {\bibfield  {journal} {\bibinfo  {journal} {J. Chem. Phys.}\ }\textbf
  {\bibinfo {volume} {131}},\ \bibinfo {pages} {104503} (\bibinfo {year}
  {2009})}\BibitemShut {NoStop}%
\bibitem [{\citenamefont {Saika-Voivod}\ \emph {et~al.}(2006)\citenamefont
  {Saika-Voivod}, \citenamefont {Poole},\ and\ \citenamefont
  {Bowles}}]{SaikaVoivod:2006p6}%
  \BibitemOpen
  \bibfield  {author} {\bibinfo {author} {\bibfnamefont {I.}~\bibnamefont
  {Saika-Voivod}}, \bibinfo {author} {\bibfnamefont {P.~H.}\ \bibnamefont
  {Poole}}, \ and\ \bibinfo {author} {\bibfnamefont {R.~K.}\ \bibnamefont
  {Bowles}},\ }\href {\doibase 10.1063/1.2203631} {\bibfield  {journal}
  {\bibinfo  {journal} {J. Chem. Phys.}\ }\textbf {\bibinfo {volume} {124}},\
  \bibinfo {pages} {224709} (\bibinfo {year} {2006})}\BibitemShut {NoStop}%
\bibitem [{\citenamefont {Frenkel}\ and\ \citenamefont
  {Smit}(2002)}]{Frenkel2002}%
  \BibitemOpen
  \bibfield  {author} {\bibinfo {author} {\bibfnamefont {D.}~\bibnamefont
  {Frenkel}}\ and\ \bibinfo {author} {\bibfnamefont {B.}~\bibnamefont {Smit}},\
  }\href@noop {} {\emph {\bibinfo {title} {{Understanding Molecular Simulation:
  From Algorithms to Applications}}}},\ edited by\ \bibinfo {editor}
  {\bibfnamefont {D.}~\bibnamefont {Frenkel}}\ and\ \bibinfo {editor}
  {\bibfnamefont {B.}~\bibnamefont {Smit}}\ (\bibinfo  {publisher} {Academic
  Press, New York},\ \bibinfo {year} {2002})\BibitemShut {NoStop}%
\bibitem [{\citenamefont {Trinkaus}(1983)}]{Trinkaus:1983eg}%
  \BibitemOpen
  \bibfield  {author} {\bibinfo {author} {\bibfnamefont {H.}~\bibnamefont
  {Trinkaus}},\ }\href {\doibase 10.1103/PhysRevB.27.7372} {\bibfield
  {journal} {\bibinfo  {journal} {Phys. Rev. B}\ }\textbf {\bibinfo {volume}
  {27}},\ \bibinfo {pages} {7372} (\bibinfo {year} {1983})}\BibitemShut
  {NoStop}%
\bibitem [{\citenamefont {Wilemski}(1999)}]{Wilemski:1999bq}%
  \BibitemOpen
  \bibfield  {author} {\bibinfo {author} {\bibfnamefont {G.}~\bibnamefont
  {Wilemski}},\ }\href {\doibase 10.1063/1.478547} {\bibfield  {journal}
  {\bibinfo  {journal} {J. Chem. Phys.}\ }\textbf {\bibinfo {volume} {110}},\
  \bibinfo {pages} {6451} (\bibinfo {year} {1999})}\BibitemShut {NoStop}%
\bibitem [{\citenamefont {Iwamatsu}(2012{\natexlab{a}})}]{Iwamatsu:2012hy}%
  \BibitemOpen
  \bibfield  {author} {\bibinfo {author} {\bibfnamefont {M.}~\bibnamefont
  {Iwamatsu}},\ }\href {\doibase 10.1063/1.3679440} {\bibfield  {journal}
  {\bibinfo  {journal} {J. Chem. Phys.}\ }\textbf {\bibinfo {volume} {136}},\
  \bibinfo {pages} {044701} (\bibinfo {year} {2012}{\natexlab{a}})}\BibitemShut
  {NoStop}%
\bibitem [{\citenamefont {Iwamatsu}(2012{\natexlab{b}})}]{Iwamatsu:2012bh}%
  \BibitemOpen
  \bibfield  {author} {\bibinfo {author} {\bibfnamefont {M.}~\bibnamefont
  {Iwamatsu}},\ }\href {\doibase 10.1063/1.4721395} {\bibfield  {journal}
  {\bibinfo  {journal} {J. Chem. Phys.}\ }\textbf {\bibinfo {volume} {136}},\
  \bibinfo {pages} {204702} (\bibinfo {year} {2012}{\natexlab{b}})}\BibitemShut
  {NoStop}%
\bibitem [{\citenamefont {Reiss}(1950)}]{Reiss:1950dd}%
  \BibitemOpen
  \bibfield  {author} {\bibinfo {author} {\bibfnamefont {H.}~\bibnamefont
  {Reiss}},\ }\href {\doibase 10.1063/1.1747784} {\bibfield  {journal}
  {\bibinfo  {journal} {J. Chem. Phys.}\ }\textbf {\bibinfo {volume} {18}},\
  \bibinfo {pages} {840} (\bibinfo {year} {1950})}\BibitemShut {NoStop}%
\bibitem [{\citenamefont {ten Wolde}\ \emph {et~al.}(1996)\citenamefont {ten
  Wolde}, \citenamefont {Ruiz-Montero},\ and\ \citenamefont
  {Frenkel}}]{ReinTenWolde:1996p1389}%
  \BibitemOpen
  \bibfield  {author} {\bibinfo {author} {\bibfnamefont {P.~R.}\ \bibnamefont
  {ten Wolde}}, \bibinfo {author} {\bibfnamefont {M.~J.}\ \bibnamefont
  {Ruiz-Montero}}, \ and\ \bibinfo {author} {\bibfnamefont {D.}~\bibnamefont
  {Frenkel}},\ }\href {\doibase 10.1063/1.471721} {\bibfield  {journal}
  {\bibinfo  {journal} {J. Chem. Phys.}\ }\textbf {\bibinfo {volume} {104}},\
  \bibinfo {pages} {9932} (\bibinfo {year} {1996})}\BibitemShut {NoStop}%
\bibitem [{\citenamefont {Shirts}\ and\ \citenamefont
  {Chodera}(2008)}]{Shirts:2008p1387}%
  \BibitemOpen
  \bibfield  {author} {\bibinfo {author} {\bibfnamefont {M.~R.}\ \bibnamefont
  {Shirts}}\ and\ \bibinfo {author} {\bibfnamefont {J.~D.}\ \bibnamefont
  {Chodera}},\ }\href {\doibase 10.1063/1.2978177} {\bibfield  {journal}
  {\bibinfo  {journal} {J. Chem. Phys.}\ }\textbf {\bibinfo {volume} {129}},\
  \bibinfo {pages} {124105} (\bibinfo {year} {2008})}\BibitemShut {NoStop}%
\bibitem [{\citenamefont {Auer}\ and\ \citenamefont
  {Frenkel}(2004)}]{Auer:2004gr}%
  \BibitemOpen
  \bibfield  {author} {\bibinfo {author} {\bibfnamefont {S.}~\bibnamefont
  {Auer}}\ and\ \bibinfo {author} {\bibfnamefont {D.}~\bibnamefont {Frenkel}},\
  }\href@noop {} {\bibfield  {journal} {\bibinfo  {journal} {J. Chem. Phys.}\
  }\textbf {\bibinfo {volume} {120}},\ \bibinfo {pages} {3015} (\bibinfo {year}
  {2004})}\BibitemShut {NoStop}%
\bibitem [{\citenamefont {Hoffman}\ \emph {et~al.}(1986)\citenamefont
  {Hoffman}, \citenamefont {Nord},\ and\ \citenamefont
  {Ruedenberg}}]{Hoffman:1986il}%
  \BibitemOpen
  \bibfield  {author} {\bibinfo {author} {\bibfnamefont {D.~K.}\ \bibnamefont
  {Hoffman}}, \bibinfo {author} {\bibfnamefont {R.~S.}\ \bibnamefont {Nord}}, \
  and\ \bibinfo {author} {\bibfnamefont {K.}~\bibnamefont {Ruedenberg}},\
  }\href {\doibase 10.1007/BF00527704} {\bibfield  {journal} {\bibinfo
  {journal} {Theoret. Chim. Acta}\ }\textbf {\bibinfo {volume} {69}},\ \bibinfo
  {pages} {265} (\bibinfo {year} {1986})}\BibitemShut {NoStop}%
\bibitem [{\citenamefont {Sear}(2005)}]{Sear:2005ca}%
  \BibitemOpen
  \bibfield  {author} {\bibinfo {author} {\bibfnamefont {R.~P.}\ \bibnamefont
  {Sear}},\ }\href {\doibase 10.1088/0953-8984/17/25/025} {\bibfield  {journal}
  {\bibinfo  {journal} {J. Phys.-Condens. Mat.}\ }\textbf {\bibinfo {volume}
  {17}},\ \bibinfo {pages} {3997} (\bibinfo {year} {2005})}\BibitemShut
  {NoStop}%
\end{thebibliography}


\end{document}